\documentclass[final]{aipproc}
\usepackage{amssymb}
\usepackage{amsmath}
\usepackage{graphicx}

\layoutstyle{6x9}

\begin{document}

\title{Black hole mass measurements using ionized gas discs: systematic dust effects}

\author{Maarten Baes}{
  address={Sterrenkundig Observatorium, Universiteit Gent, Krijgslaan 281 S9, B-9000 Gent, Belgium} 
}

\begin{abstract}
  Using detailed Monte Carlo radiative transfer simulations in
  realistic models for galactic nuclei, we investigate the influence
  of interstellar dust in ionized gas discs on the rotation curves and
  the resulting black hole mass measurements. We find that absorption
  and scattering by interstellar dust leaves the shape of the rotation
  curves basically unaltered, but slightly decreases the central slope
  of the rotation curves. As a result, the ``observed'' black hole
  masses are systematically underestimated by some 10 to 20\% for
  realistic optical depths. We therefore argue that the systematic
  effect of dust attenuation should be taken into account when
  estimating SMBH masses using ionized gas kinematics.
\end{abstract}

\maketitle


\section{Introduction}

Measuring the kinematics of ionized gas disks has become one the most
important methods to determine the masses of supermassive black holes
(SMBHs) in the nuclei of nearby galaxies. The state-of-the-art
modeling techniques are refined to a high degree and take into account
the major relevant astrophysical processes as well as the main
instrumental effects. Applying such modeling techniques to
high-quality HST data, a formal SMBH mass measurement uncertainty of
25\% or better has been claimed with well-behaved disks
(e.g. \cite{Bar03}, \cite{Mar03}, \cite{Atk05}, \cite{Har06}). Many
ionized gas disks contain substantial amounts of interstellar dust. As
optical H$\alpha$ radiation is easily absorbed and scattered by
interstellar dust grains, these effects might affect the observed
rotation curve and hence the SMBH mass estimate. The goal of our
investigation is to determine the importance of this effect through
detailed Monte Carlo simulations.

\section{The model}

We have set up a detailed radiative transfer model to investigate the
effects of neglecting dust absorption and scattering on ionized gas
disc rotation curves and the corresponding SMBH masses in a typical
galactic nucleus. Our model consists of a thin (but not
infinitesimally thin) axisymmetric double-exponential disc rotating in
a galactic nucleus. The intrinsic rotation velocity of the disc is
determined by the combined gravitational potential of a stellar
distribution and a central SMBH. We consider various SMBH masses, up
to $M_\bullet=10^8~M_\odot$. An additional isotropic bulk velocity
dispersion is taken into account. We assume that a fraction of the
disc consists of interstellar dust with typical Milky Way dust optical
properties.  The dust is assumed to be coupled to the ionized gas and
has exactly the same spatial and velocity distribution. The SMBH mass
$M_\bullet$, the thickness $z_0$ of the disc and the V-band optical
depth $\tau_{\text{\tiny{V}}}$, a quantity equivalent to the dust
mass, are free parameters in our models.

\section{SKIRT radiative transfer modelling}

For each model, we calculate the observed line-of-sight velocity field
(data cubes) at several inclination angles. We use the SKIRT code, a
3D Monte Carlo radiative transfer code that has been developed
explicitly to model the kinematics of dusty galaxies \cite{Bae02},
\cite{Bae03}. The code uses photon packages that carry around
information on the line-of-sight velocity of the sources (in this case
the H$\alpha$ emitting gas) that emitted them. The effects of
absorption and multiple anisotropic scattering are properly taken into
account. During scattering events, the Doppler information carried by
each photon package is updated, taking into account both the original
line-of-sight velocity and the scattering dust grain velocity. When
the photon packages leave the system, their ``observed'' line-of-sight
velocity information can be extracted, which is used to build up data
cubes. At the end of the simulation, we calculate rotation velocity
maps and velocity dispersion maps from the simulated data cubes. We
focus in particular on the major axis rotation curves.

\section{Results}

\begin{figure}
\centering
\includegraphics[width=\textwidth]{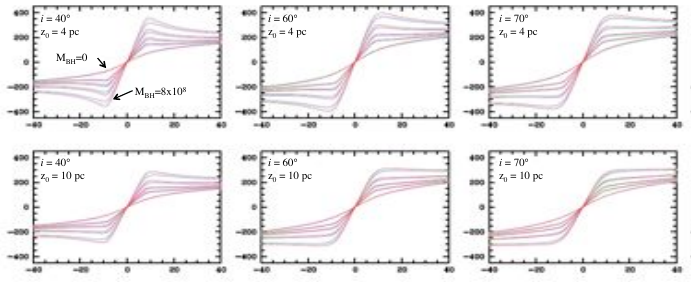}
\caption{Nuclear rotation curves for dusty ionized gas discs. Each
panel corresponds to a particular choice for the inclination $i$ and
the disc scale height $z_0$. In each panel we plot the rotation curves
for different black hole masses (ranging between 0 and
$8\times10^8$~M$_\odot$) and for different optical depths (ranging
between 0 and 2). For any $i$, $z_0$ and $M_\bullet$, the slope of the
rotation curve slightly decreases with increasing optical depth.}
\label{mbaesfig1}
\end{figure}

In Figure~{\ref{mbaesfig1}} we plot a set of nuclear rotation curves
for our dusty ionized gas discs. The different panels correspond to
different values for $z_0$ and the inclination angle. Within each
individual panel, rotation curves are shown for different values of
$M_\bullet$ and $\tau_{\text{\tiny{V}}}$. For a fixed spatial
resolution and inclination, the shape of the rotation curve of
dust-free models depends on the SMBH mass and disc thickness. In
particular, increasing disc thickness has a significant influence due
to the larger integration along the line of sight. Modeling ionized
gas disks as infinitesimally thin disks might introduce a bias.

\begin{figure*}
  \centering
  \includegraphics[width=\textwidth]{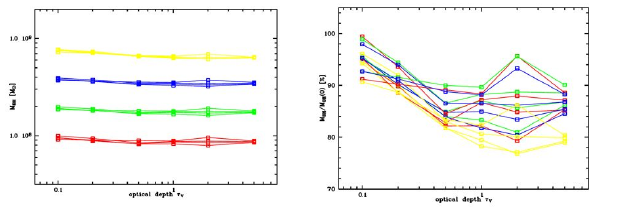}
  \caption{{\bf{Left panel:}} An estimate of the ``observed'' black
    hole mass of the galaxy, derived from the observed dust-affected
    rotation curve, explicitly as a function of optical depth. The
    different colours correspond to different intrinsic SMBH masses,
    the different lines correspond to different inclination angles.
    --- {\bf{Right panel:}} The ratio between the ``observed'' and the
    intrinsic SMBH mass, explicitly as a function of optical
    depth. SMBH masses are systematically underestimated if dust
    attenuation is not properly taken into account.}
  \label{mbaesfig2}
\end{figure*}

To investigate the effect of interstellar dust absorption and
scattering, we must compare models with different values of the
optical depth. The curves corresponding to various
$\tau_{\text{\tiny{V}}}$ basically coincide: apparently absorption and
scattering by interstellar dust leaves the shape of the rotation
curves qualitatively unaltered. However, close inspection shows that
an increasing amount of interstellar dust slightly decreases the
central slope of the rotation curves. We estimate the bias one makes
by neglecting the effect of interstellar dust on SMBH mass estimates
by translating the observed rotation curve central slopes into SMBH
masses. Figure~{\ref{mbaesfig2}} shows the effect of increasing dust
mass on the ``observed'' SMBH mass and on the ratio between the
``observed'' and the intrinsic SMBH mass. It clearly demonstrates
that, as the optical depth increases, the observed SMBH mass
decreases, i.e. the true SMBH mass is more and more
underestimated. This means that neglecting interstellar dust when
determining SMBH masses from ionized gas kinematics systematically
leads to an underestimate of the black hole mass. For
$\tau_{\text{\tiny{V}}} \sim 0.2$ the SMBH is underestimated by about
10\%, for $\tau_{\text{\tiny{V}}} > 1$ the black hole mass is
underestimated by 10 to 20\%. This effect is more or less independent
of the SMBH mass and the inclination of the disc.

\section{Conclusions}

We find that absorption and scattering by interstellar dust leaves the
shape of the rotation curves basically unaltered, but slightly
decreases the central slope of the rotation curves. Since the central
slope can be used as a proxy for the SMBH mass, this result means that
``observed'' black hole masses are systematically underestimated by
some 10 to 20\% for realistic optical depths. Our study demonstrates
that the systematic effect of dust attenuation should be taken into
account in SMBH demographics studies.

\end{document}